\documentclass[prd,onecolumn,english]{revtex4}
\usepackage{graphicx}
\begin{document}

\large

\title{Electroweak and QCD  Radiative Corrections 
to Drell--Yan Process for Experiments at the Large Hadron Collider}

\author{Vladimir A. Zykunov}
\email{vladimir.zykunov@cern.ch, zykunov@rambler.ru}
\affiliation{Belarussian State University of Transport,  246653 Gomel, Belarus}

\begin{abstract}
Next-to-leading order electroweak and QCD radiative corrections
to the Drell--Yan process with high dimuon masses  
for experiments CMS LHC at CERN have been studied in fully differential form.
The FORTRAN code READY for numerical analysis of Drell--Yan observables has been presented.
The radiative corrections are found to become significant for CMS LHC experiment setup.
\end{abstract}

\maketitle

\section{Introduction}

For more than twenty years the Standard Model (SM) has had the status
of a consistent and experimentally confirmed theory since the experimental
data of past and present accelerators (LEP, SLC, Tevatron)
has shown no significant deviation from SM predictions up to the energy
scale of a few hundred GeV and, finally, LHC has discovered Higgs boson	\cite{new-boson}. 
However, various New Physics (NP) models such as 
production of high-mass dilepton resonances \cite{extra-bos}, 
extra spatial dimensions \cite{extra-dim} etc.
suggest deviations beyond SM predictions and testing them at the new energy scale (the few thousand GeV region) 
is one of the main tasks of modern physics. 
The forthcoming experiments at the LHC with maximal energy
would either provide the first data on NP or strengthen the current status of the SM.

The experimental investigation of the continuum for the Drell-Yan
production of dileptons, i.e. data on the cross section and the forward-backward
asymmetry of the reaction 
\begin{equation}
pp\rightarrow(\gamma,Z)\rightarrow l^{+}l^{-}X\label{1}
\end{equation}
at large invariant mass of a dilepton pair (see \cite{cmsnote} and references therein) 
is considered to be one of the most powerful tools
in the experiments at the LHC from a NP exploration standpoint.

The studies of the NP effects are impossible without exact knowledge
of the SM predictions including higher-order electroweak (EWK) and QCD radiative corrections. 
Many programs have been developed for this: 
DYNNLO, FEWZ, HORACE, MC@NLO, POWHEG, RADY, READY, SANC, ZGRAD/ZGRAD2 et al.
A large list of references quoted, for example, in recent papers \cite{FEWZ,POWHEG}
dedicated to description of FEWZ and POWHEG, correspondingly.
These codes were used for taking into account the uncertainty due to the EWK and QCD  corrections
at recent measurements of the differential ${d\sigma}/{dM}$  ($M$ is dilepton invariant mass)
and double-differential ${d^2\sigma}/{(dM dy)}$ ($y$ is dilepton rapidity)
Drell--Yan cross sections  at  LHC energy $\sqrt{S}$ = 7 TeV, $M \leq 1.5 \ \mbox{TeV}$
and integrated luminosity 4.5 $\mbox{fb}^{-1}$ \cite{CMS-PAS-EWK-11-007}.
Measurements are in agreement with the SM predictions:
all of them  with next-to-next-to-leading order (NNLO) of FEWZ using MSTW2008 parton density functions (PDF)
and double-differential observables with NLO of POWHEG using CT10 PDF.

At the edges of kinematical region (especially at extra large $M$)
the important task is to make the correction procedure of background both accurate and fast. 
For the latter it is desirable to obtain the set of 
as much compact as possible formulas both for EWK and QCD corrections.
{ To get leading effect of weak corrections in the region of large
invariant dilepton mass we actively used the so-called Sudakov logarithms (SL)
\cite{sud-log} which grow with the energy scale and thus give one
of the main effects in the region of large invariant dilepton mass.
In addition, the collinear logarithms (CL) of the QED and QCD radiative corrections 
can compete with double SL in the investigated region.}
Such formulas have been obtained in previous papers 
\cite{YAFDY}--\cite{qcd2}
using the asymptotic approach for the most complicated weak components of EWK corrections, 
and using the leading CL extraction \cite{LL,qcd1,qcd2} for the QED and QCD component. 
This paper is devoted to the analysis of the interplay of these effects for 
observable quantities of CMS LHC in general fully differential form.

\section{Notations and cross sections with the Born kinematics}

At LO the Drell--Yan process  $p(P_{A})+p(P_{B}) \rightarrow l^{+}(k_{1})+l^{-}(k_{2})+X$
in quark-parton model is described by Fig.1,a.
Our notations are the following: 
$p_{1}(p_{2})$ is the 4-momentum of the quark or antiquark with flavor $q$ and mass
$m_{q}$ from the incoming proton with 4-momentum $P_{A}$ or $P_{B}$;
\ $k_{1}(k_{2})$ is the 4-momentum of the final lepton $l^{+}(l^{-})$
with mass $m$;\ \
$q=k_{1}+k_{2}$ is the 4-momentum of the $i$-boson with mass $m_{i}$
($i=\gamma,Z;\ m_{\gamma}=0$). We use the standard set of Mandelstam
invariants for the partonic elastic scattering: 
\begin{equation}
s=(p_{1}+p_{2})^{2},\ t=(p_{1}-k_{1})^{2},\ u=(k_{1}-p_{2})^{2},
\end{equation}
 and $S=(P_{A}+P_{B})^{2}$ for hadron scattering. 
The invariant mass of the dilepton is $M=\sqrt{q^{2}}$.

\begin{figure*}
\vspace*{-10mm} 
\scalebox{0.17}{\includegraphics{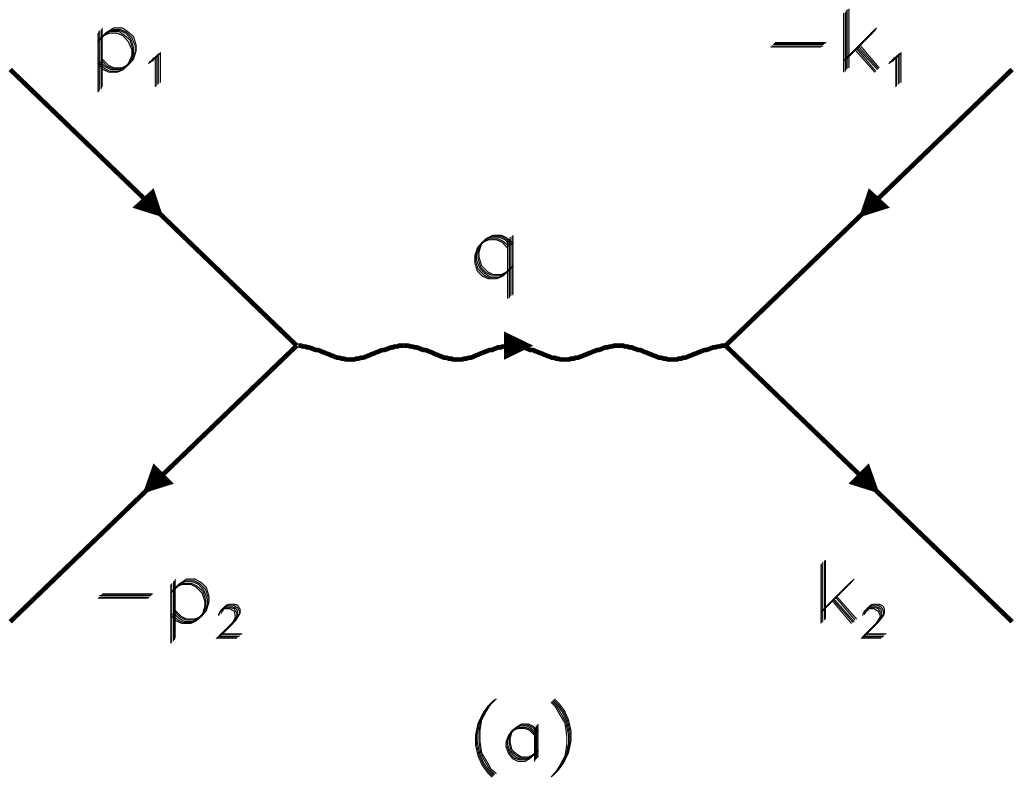}} \hspace*{-11mm}
\scalebox{0.17}{\includegraphics{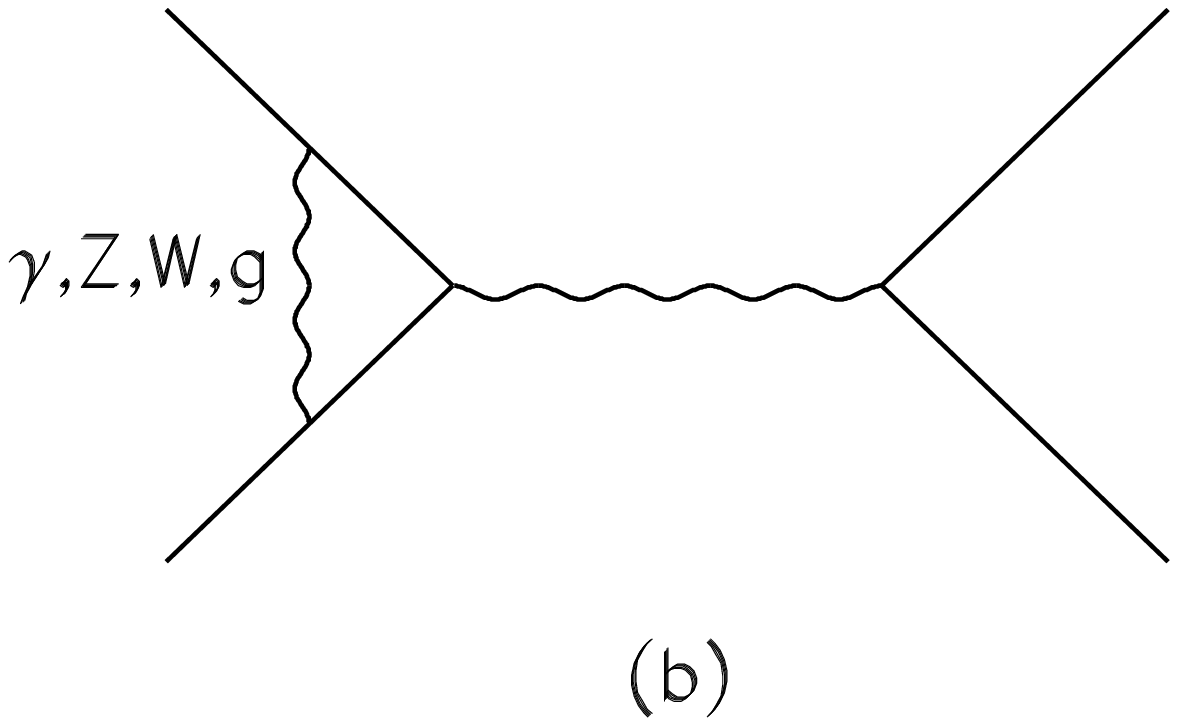}}  \hspace*{-11mm}
\scalebox{0.17}{\includegraphics{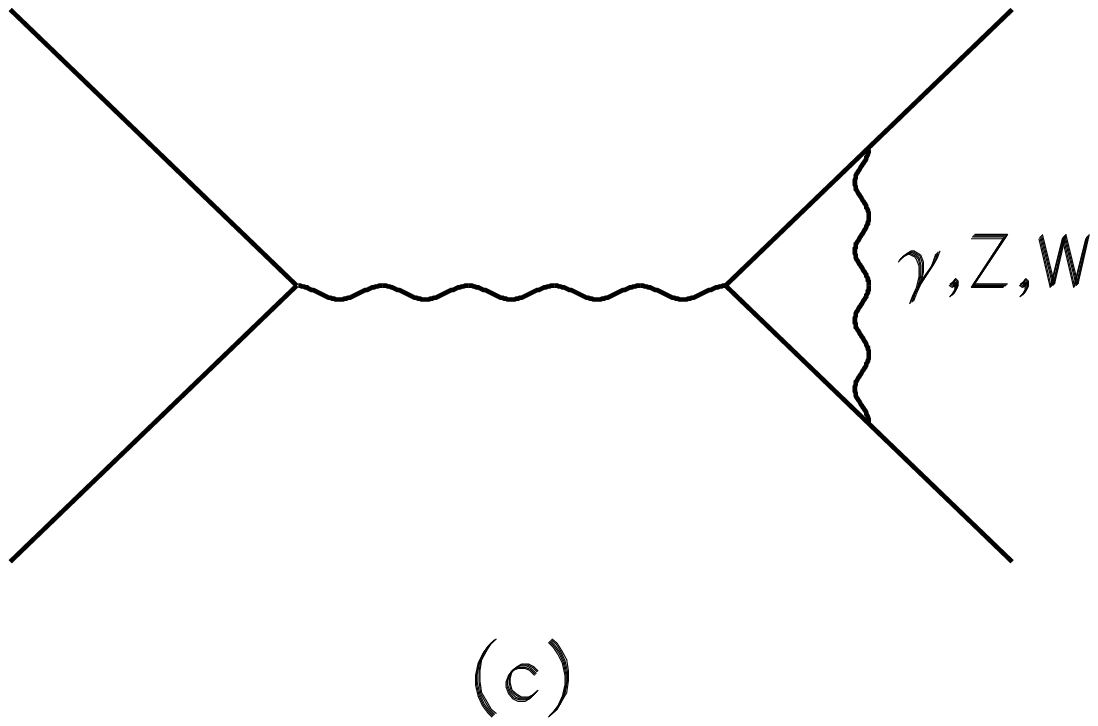}}  \hspace*{-11mm}
\scalebox{0.17}{\includegraphics{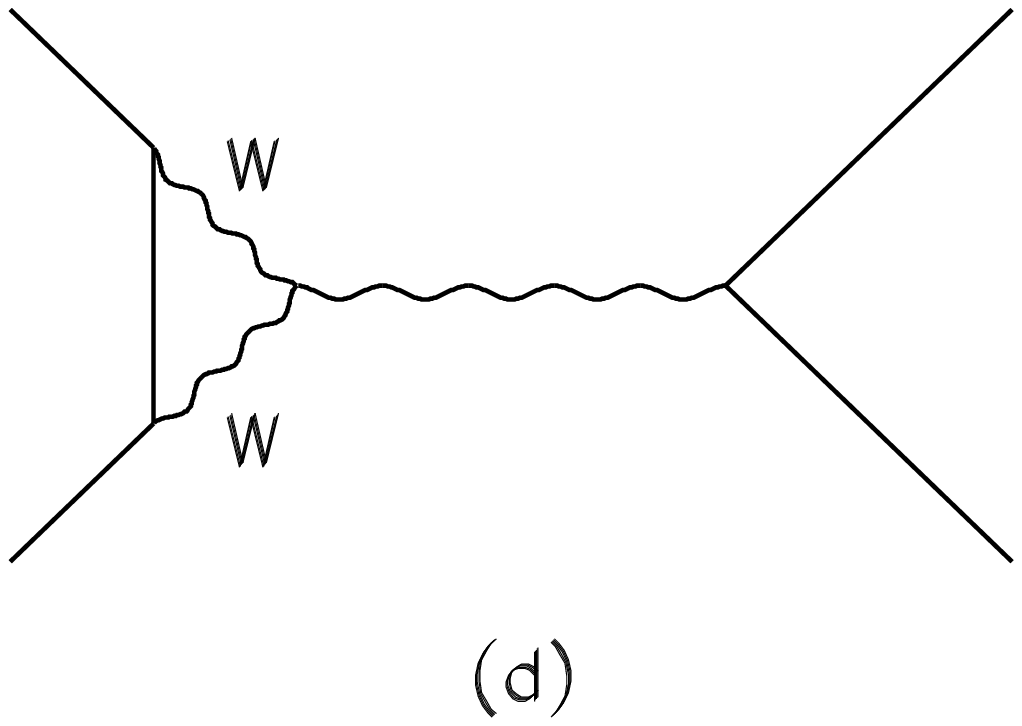}}  \hspace*{-11mm}
\scalebox{0.17}{\includegraphics{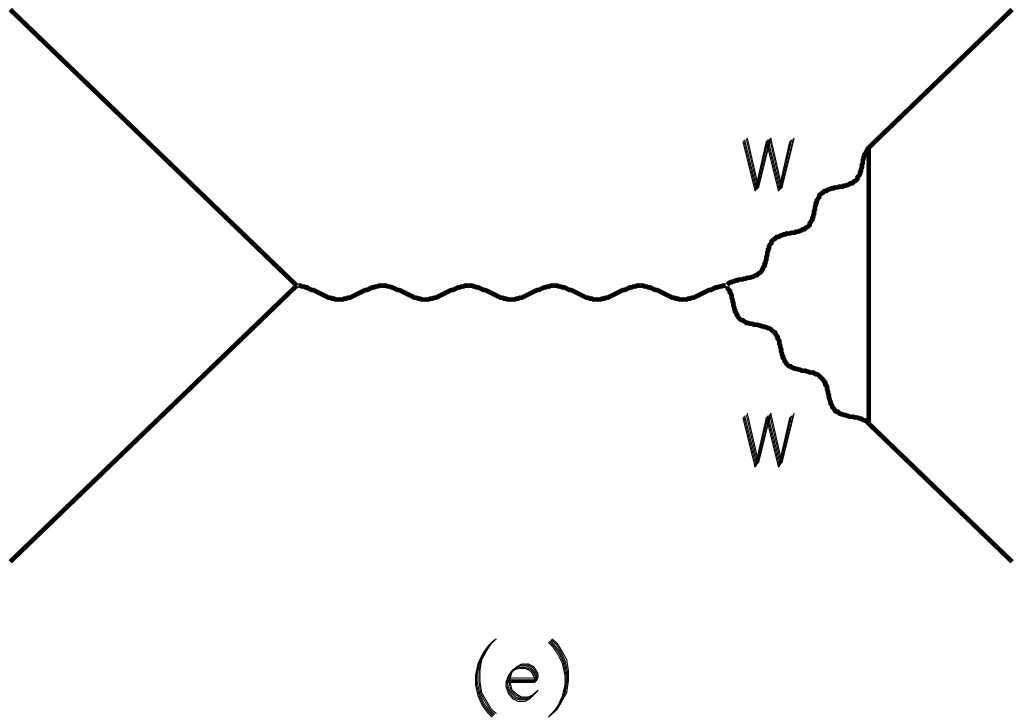}}  \hspace*{-11mm}
\scalebox{0.17}{\includegraphics{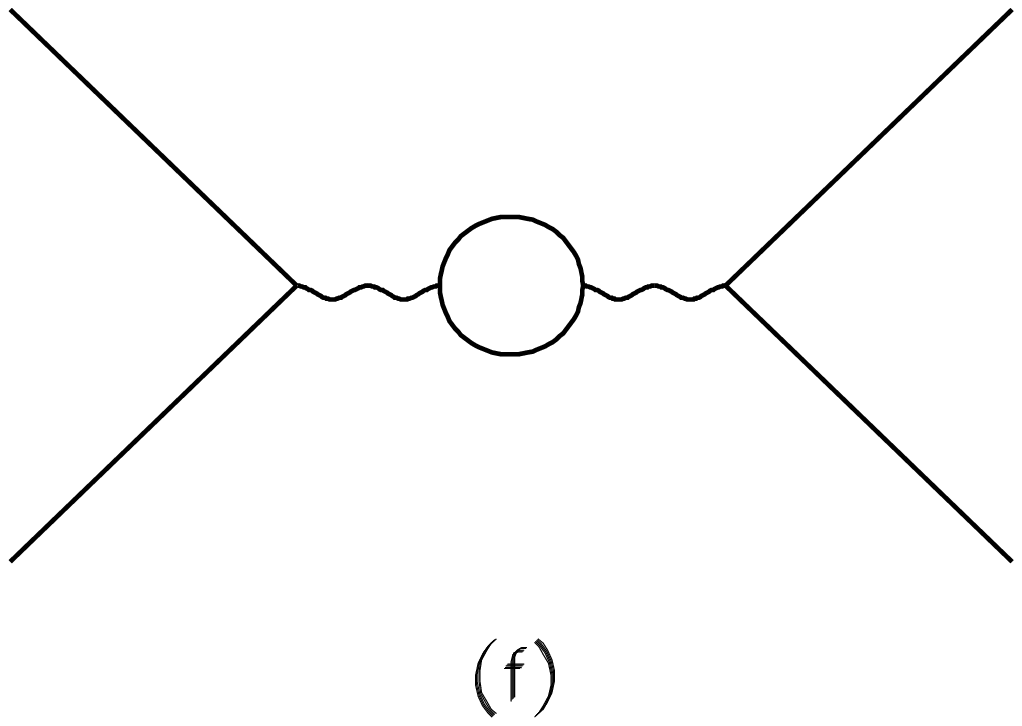}}  
\vspace{-15mm} 
\newline
\hspace*{-06mm}
\scalebox{0.16}{\includegraphics{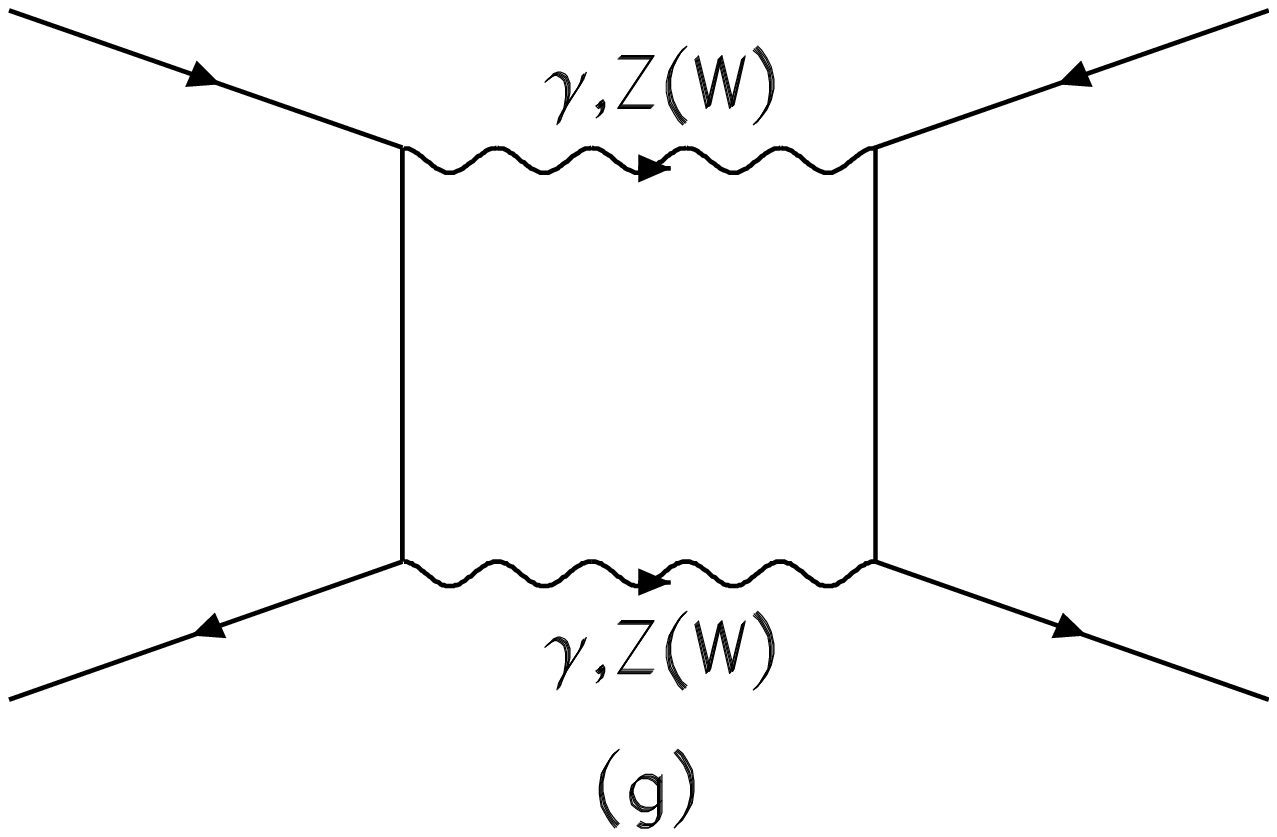}} \hspace*{-09mm}
\scalebox{0.16}{\includegraphics{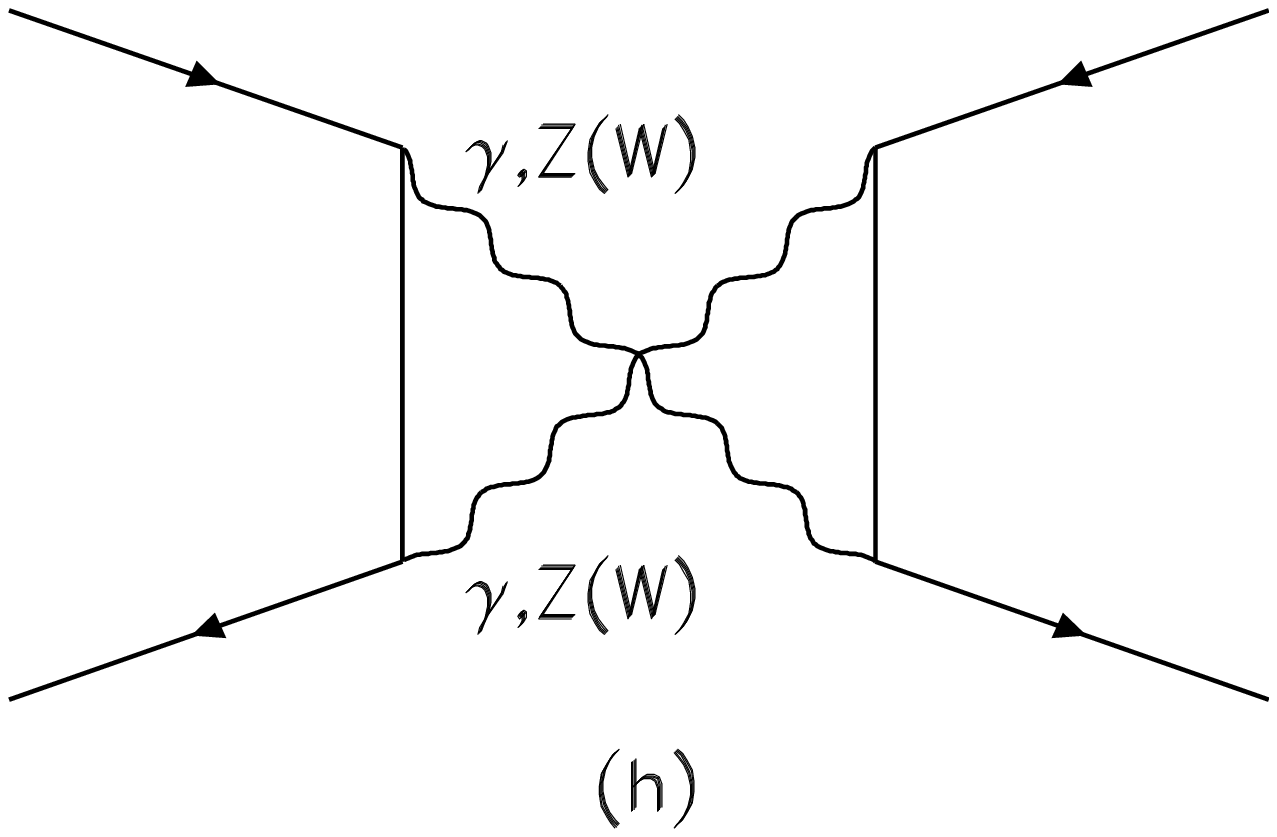}} \hspace*{-06mm}
\scalebox{0.17}{\includegraphics{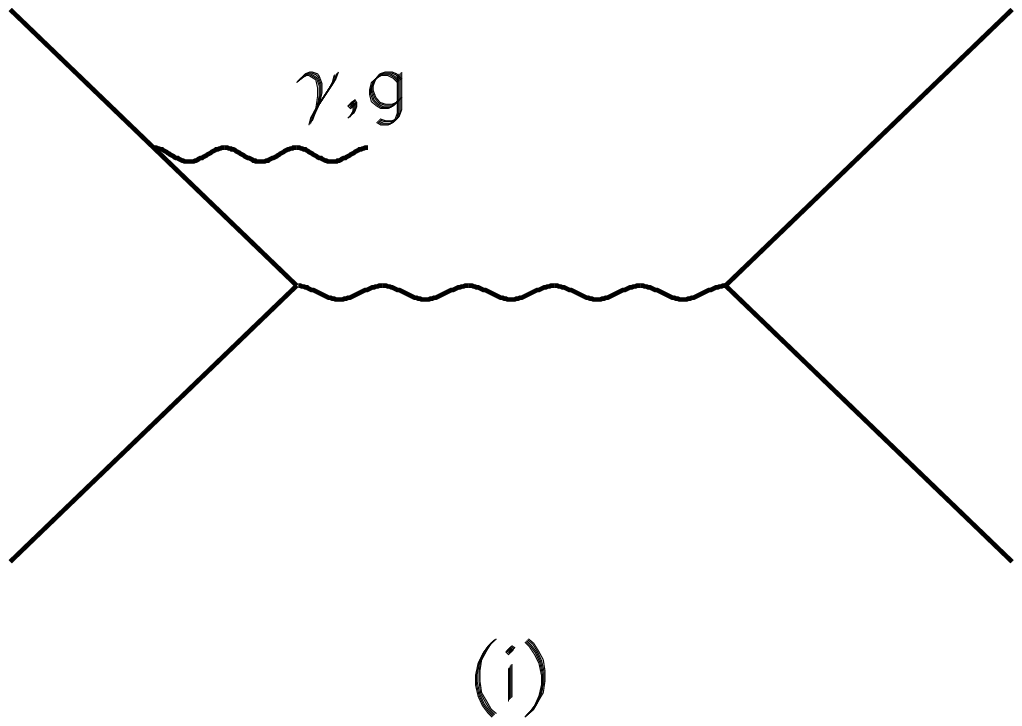}} \hspace*{-11mm}
\scalebox{0.17}{\includegraphics{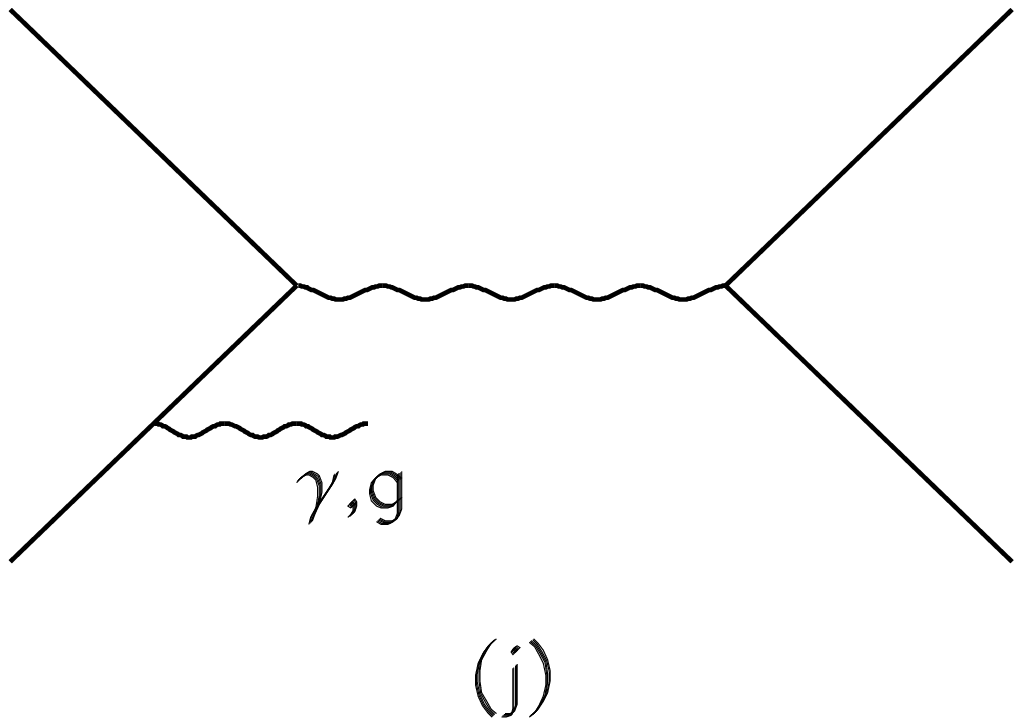}} \hspace*{-11mm}
\scalebox{0.17}{\includegraphics{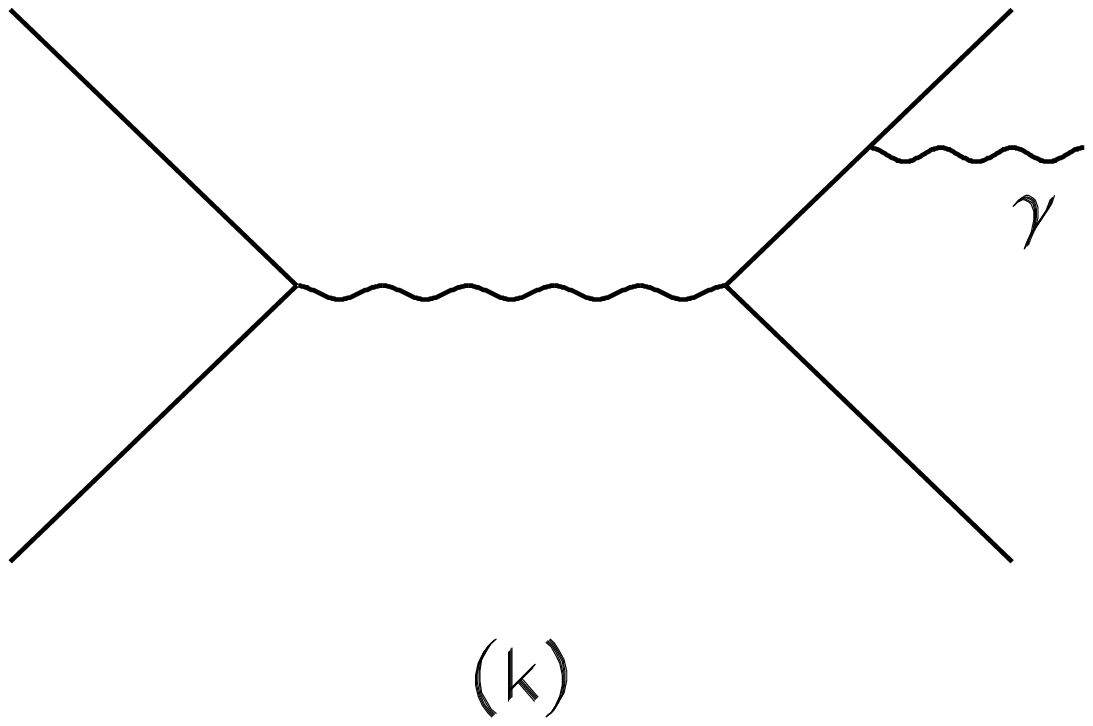}} \hspace*{-11mm}
\scalebox{0.17}{\includegraphics{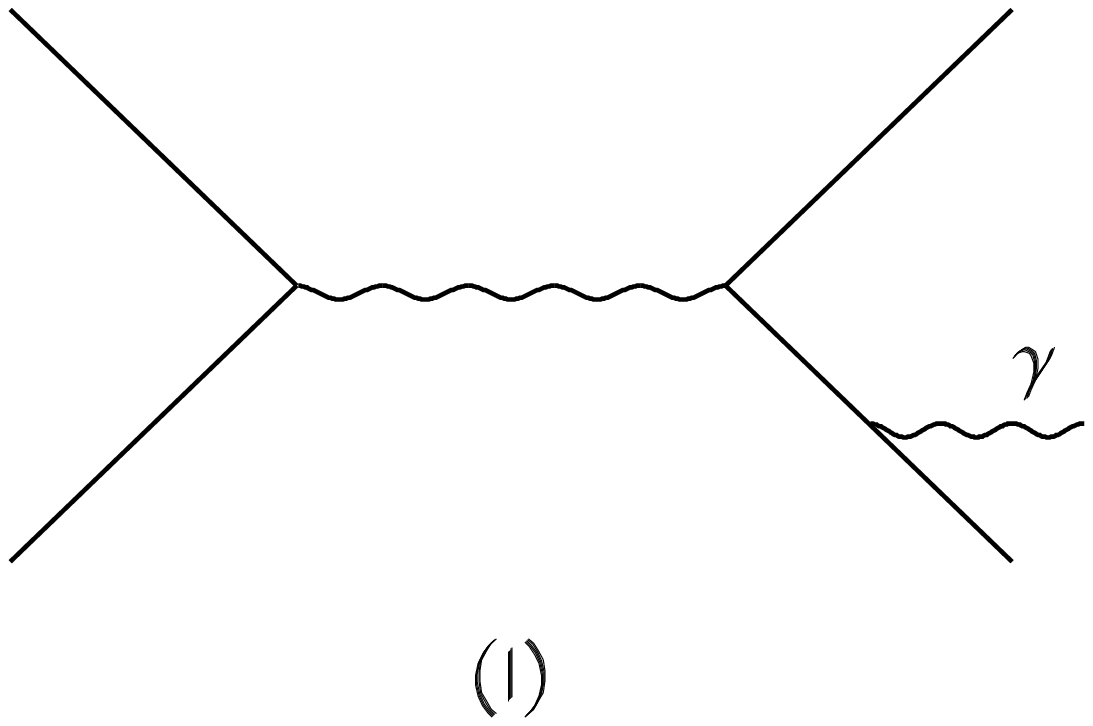}} 
\vspace*{-4mm}
 \caption{\label{fig:1} Feynman graphs for the Born (a), one-loop virtual diagrams
(b-h), and bremsstrahlung diagrams for the Initial (ISR) (i, j) and
Final State Radiation (FSR) (k, l). The unlabeled wavy lines stand
for the virtual $\gamma$ or $Z$ boson. }
\vspace*{2mm}
\end{figure*}

Let us start by presenting the convolution formula for the total cross
section with non-radiative kinematics: 
\begin{eqnarray}
\sigma_{N}=\frac{1}{3}\int\limits _{0}^{1}dx_{1}\int\limits _{0}^{1}dx_{2}
\int\limits _{-S}^{0}dt  \sum\limits _{q=u,d,...}  
 && \!\!\!\!\!\!\!\!\!\!\! \Bigl[ f_{q}^{A}(x_{1},Q^{2})f_{\bar{q}}^{B}(x_{2},Q^{2}) \sigma_{N}^{q\bar{q}}(t)+
\nonumber \\ [0.1cm]
\displaystyle 
 && \!\!\!\!\!\!\!\!\!\!\!\!\!\!  +  f_{\bar{q}}^{A}(x_{1},Q^{2})f_{q}^{B}(x_{2},Q^{2}) \sigma_{N}^{\bar{q}q}(t) \Bigr]
\theta(s+t)\theta_{M}\theta_{D}.
\label{xsfin}
\end{eqnarray}
Here, $f_{q}^{H}(x,Q^{2})dx$ is the probability of finding, in hadron
$H$, a quark $q$ at energy scale~$Q^{2}$ carrying a momentum fraction
between $x$ and $x+dx$, and $\sigma^{q\bar{q}}$ and $\sigma^{\bar{q}q}$
are the cross sections at the quark-parton level. According to the
quark-parton model rules, we take $s=x_{1}x_{2}S$. 
The function 
$\theta_{M}=\theta(s-M_{1}^{2})\theta(M_{2}^{2}-s)$
 under the integral sign is determined by the kinematics of the parton
reaction and provides the integration in the interval of invariant
mass $M_{1}\leq M\leq M_{2}$. The factor 
\begin{equation}
\theta_{D}=\theta(\zeta^{*}-\zeta)\theta(\zeta^{*}+\zeta)\theta(\zeta^{*}-\psi)\theta(\zeta^{*}+\psi)
\theta(p_{T}(l^{+})-p_{T}^{\rm min})\theta(p_{T}(l^{-})-p_{T}^{\rm min})
\label{tetad}
\end{equation}
 cuts the region of integration according to detector geometry. Here,
$\zeta=\cos\theta_{1},\ \psi=\cos\theta_{2}$, where $\theta_{1(2)}$
is the scattering angle of the lepton with 4-momenta $k_{1(2)}$ in
the hadron center-of-mass frame. For the CMS detector the parameter
$\zeta^{*}\approx0.986614$ corresponds to the lepton rapidity limitation
$y(l)^{*}=2.5$. For the transverse components of lepton momenta we
have the relations $p_{T}(l^{+})={k_{1}}_{0}\sin\theta_{1}$ and $p_{T}(l^{-})={k_{2}}_{0}\sin\theta_{2}$,
and, for the CMS detector, $p_{T}^{\rm min}=20\ \mbox{GeV}$.

We use the common index for contributions with non-radiative kinematics $N=\{ 0,V,{\rm soft} \}$,
where 0 stands for the Born contribution,
the special indices for contributions with at least one additional virtual particle $V=\{ \mbox{BSE},\mbox{HV},b \}$
and separately for the contributions of boxes $b=\{ \gamma\gamma,\gamma Z,ZZ,WW \}$.
Abbreviations mean: 
BSE for boson self energies, 
HV for Heavy Vertices induced by at least one massive boson, 
"$\gamma\gamma$" for the infrared(IR)-finite part of $\gamma\gamma$-boxes, 
"$\gamma Z$" for the IR-finite part of $\gamma Z$-boxes 
and "$ZZ$" for the $ZZ$-boxes, "$WW$" for the $WW$-boxes,
"$\rm soft$" for the sum of Light Vertices (LV) induced by one massless photon or gluon, 
the IR-divergent parts of the $\gamma\gamma$-boxes, $\gamma Z$-boxes, and of the (soft) bremsstrahlung cross section. 
The "$\rm soft$"-part is IR-finite in sum and described by Born kinematics.

Now, let us present explicit formulae for $q\bar q$ cross sections
given in (\ref{xsfin}), employing the notation $\sigma(t)\equiv d\sigma/{dt}$.
To find the cross section for the ${\bar{q} q}$-case, we can use crossing rules.
The Born cross section has the form 
\begin{eqnarray}
\sigma_{0}^{q\bar{q}}(t)
=\frac{2\pi\alpha^{2}}{s^{2}}\sum\limits _{i,j=\gamma,Z}D^{i}{D^{j}}^{*}
\sum\limits _{\chi=+,-}{\lambda_{q}}_{\chi}^{i,j}{\lambda_{l}}_{\chi}^{i,j}B_{\chi},
\end{eqnarray}
 where the boson propagators look like 
$D^{j}=({s-m_{j}^{2}+im_{j}\Gamma_{j}})^{-1}$,
 $\Gamma_{j}$ is the $j$-boson width, $B_{\chi}=t^{2}+\chi u^{2}$. 
%
The combinations of coupling constants for a $f$-fermion with
an $i$- (or $j$-) boson have the form \begin{equation}
{\lambda_{f}}_{+}^{i,j}=v_{f}^{i}v_{f}^{j}+a_{f}^{i}a_{f}^{j},\ {\lambda_{f}}_{-}^{i,j}=v_{f}^{i}a_{f}^{j}+a_{f}^{i}v_{f}^{j},\label{lamb}\end{equation}
 where 
\begin{equation}
v_{f}^{\gamma}=-Q_{f},\ a_{f}^{\gamma}=0,\ v_{f}^{Z}=\frac{I_{f}^{3}-2s_{W}^{2}Q_{f}}{2s_{W}c_{W}},\ a_{f}^{Z}=\frac{I_{f}^{3}}{2s_{W}c_{W}},
\end{equation}
 $Q_{f}$ is the electric charge of fermion $f$ in proton charge
units $e,\ (e=\sqrt{4\pi\alpha})$, $I_{f}^{3}$ is the third component
of the weak isospin of fermion $f$, and $s_{W}\ (c_{W})$ is the
sine(cosine) of the weak mixing angle. 

The BSE-part is 
\begin{eqnarray}
\sigma_{{\rm BSE}}^{q\bar{q}}(t)=
-\frac{4\alpha^{2}\pi}{s^{2}} \Bigl[ 
 && \!\!\!\!\!\!\!\!\!\!\! \sum\limits _{i,j=\gamma,Z}\Pi_{S}^{i}D^{i}{D^{j}}^{*}
\sum\limits _{\chi=+,-}{\lambda_{q}}_{\chi}^{i,j}{\lambda_{l}}_{\chi}^{i,j}B_{\chi}+
\nonumber \\
 && \!\!\!\!\!\!\!\!\!\!\! + 
\Pi_{S}^{\gamma Z}D^{Z}\sum\limits _{i=\gamma,Z}{D^{j}}^{*} \sum\limits _{\chi=+,-}
({\lambda_{q}}_{\chi}^{\gamma,j}{\lambda_{l}}_{\chi}^{Z,j}+{\lambda_{q}}_{\chi}^{Z,j}{\lambda_{l}}_{\chi}^{\gamma,j})B_{\chi}
 \Bigr].
\end{eqnarray}
 Here $\Pi_{S}^{\gamma,Z,\gamma Z}$
are connected with the renormalized photon--, $Z$-- and $\gamma Z$--self
energies \cite{BSH86,Hollik} as 
\[
\Pi_{S}^{\gamma}=\frac{\hat{\Sigma}^{\gamma}}{s},\ 
\Pi_{S}^{Z}=\frac{\hat{\Sigma}^{Z}}{s-m_{Z}^{2}},\ 
\Pi_{S}^{\gamma Z}=\frac{\hat{\Sigma}^{\gamma Z}}{s}.\]
The HV-part has the following form: 
\begin{eqnarray}
\sigma_{{\rm HV}}^{q\bar{q}}(t)=\frac{4\pi\alpha^{2}}{s^{2}}{\rm Re}\sum_{i,j=\gamma,Z}D^{i}{D^{j}}^{*}\sum_{\chi=+,-}
({\lambda_{q}^{{\rm F}}}_{\chi}^{i,j}{\lambda_{l}}_{\chi}^{i,j}+{\lambda_{q}}_{\chi}^{i,j}{\lambda_{l}^{{\rm F}}}_{\chi}^{i,j})B_{\chi},
\end{eqnarray}
 where the form factors ${\lambda_{f}^{{\rm F}}}_{\pm}^{i,j}$ are
given in \cite{YAFDY}.
The boxes can be presented as 
\begin{eqnarray}
\sigma_{b}^{q\bar{q}}(t)=\frac{2\alpha^{3}}{s^{2}}
\sum_{k=\gamma,Z}{D^{k}}^{*} \bigl[ \delta^{b,k}(t,u,b_{+},b_{-})-\delta^{b,k}(u,t,b_{-},b_{+}) \bigr],
\end{eqnarray}
where the functions $\delta^{b,k}(t,u,b_{+},b_{-})$, $b_{\chi}$
and all prescriptions for them can be found in \cite{YAFDY,PRD}.

The QED "$\rm soft$"-part 
(the result of infrared singularity cancellation of $\gamma\gamma,\ \gamma Z,\ \mbox{photon LV}$ 
and soft photon bremsstrahlung) is proportional to Born cross section:
\begin{eqnarray}
\sigma_{\rm soft}^{q\bar{q}}(t)= \frac{\alpha}{\pi} \delta_{\rm soft}^{q\bar{q}}\sigma_{0}^{q\bar{q}}(t)
\label{soft-xs}
\end{eqnarray}
with corresponfing factor
\begin{eqnarray}
\delta_{\rm soft}^{q\bar{q}}= 
&& \!\!\!\!\!\!\!\!\!	
2 \ln\frac{2\omega}{\sqrt{s}}
\Bigl[ 
Q_{q}^{2}\bigl(\ln\frac{s}{m_{q}^{2}}-1\bigr) - 2Q_{q}Q_{l}\ln\frac{t}{u} + Q_{l}^{2}\bigl(\ln\frac{s}{m^{2}}-1\bigr)
\Bigr] 
+Q_{q}^{2} \Bigl(\frac{3}{2}\ln\frac{s}{m_{q}^{2}}-2+\frac{\pi^{2}}{3}\Bigr) -
\nonumber \\[0.3cm]
{\displaystyle } 
&& \!\!\!\!\!\!\!\!\! 
-Q_{q}Q_{l}
\Bigl(\ln\frac{s^{2}}{tu}\ln\frac{t}{u}+\frac{\pi^{2}}{3}+\ln^{2}\frac{t}{u}+4{\mbox{ {\rm Li}}}_{2}\frac{-t}{u}\Bigr)
+Q_{l}^{2} \Bigl(\frac{3}{2}\ln\frac{s}{m^{2}}-2+\frac{\pi^{2}}{3}\Bigr),
\label{soft}
\end{eqnarray}
 where $\omega$ is a parameter that determines the "softness"
of a photon -- the maximal energy of a soft photon, and $\mbox{{\rm Li}}_{2}$
denotes the Spence dilogarithm. 
The QCD "$\rm soft$"-part can be found from (\ref{soft}) by neglecting the FSR
and interference parts and after substitution:
\begin{equation}
C_{\rm QED} = Q_q^2  \frac{\alpha}{\pi} 
\rightarrow  \sum\limits_{a=1}^{N^2-1}  t^at^a \frac{\alpha_s}{\pi}  
= \frac{N^2-1}{2N} I \frac{\alpha_s}{\pi}
  \rightarrow \frac{4}{3} \frac{\alpha_s}{\pi} = C_{\rm QCD},
\label{zamena}
\end{equation}
where  $N=3$, and $2 t^a$ are Gell-Mann matrices.

\section{Hard photons and gluons. Inverse gluon emission}

Let us present the Drell--Yan cross section contribution induced by bremsstrahlung (Fig.1(i-l)). 
We introduce the total phase space of the reaction 
$ p(P_{A})+p(P_{B}) \rightarrow l^{+}(k_{1})+l^{-}(k_{2})+b(k)+X,\  (b=\gamma \ \mbox{or}\ g) $  as 
\begin{eqnarray}
I_{\Omega}[A]=\int\limits _{0}^{1}dx_{1}\int\limits _{0}^{1}dx_{2}
\int\!\!\!\!\int\limits _{~\Omega}\!\!\!\!\int\!\!\!\!\int 
dtdvdzdu_{1}\frac{1}{\pi\sqrt{R_{u_{1}}}}\theta(R_{u_{1}})\theta_{M}^{R}\theta_{D}^{R}\ A,
\label{i6}
\end{eqnarray}
where $z=2k_{1}k,\ v=2k_{2}k,\ z_{1}=2p_{1}k,\ u_{1}=2p_{2}k$ (for radiative kinematics $v=s+t+u$ and $z+v = z_{1}+u_{1}$)
and
$k$ is the 4-momentum of a real bremsstrahlung photon (gluon).

The factor $\theta_{M}^{R}$ for the radiative case has the form 
\begin{equation}
\theta_{M}^{R}=\theta(s-z-v-M_{1}^{2})\theta(M_{2}^{2}-s+z+v).
\end{equation}
 For $\theta_{D}^{R}$, we use the "non-radiative"\ expression
$\theta_{D}$ (\ref{tetad}) 
with the angles and energies depending on additional "radiative"
invariants: \begin{eqnarray}
 &  & \zeta=\frac{x_{1}u-x_{2}t}{x_{1}u+x_{2}t},\  
      \psi= \frac{x_{1}(s+u-u_{1})+x_{2}(u+z_{1}-v)}{x_{1}(s+u-u_{1})-x_{2}(u+z_{1}-v)},
\label{psss}\\[0.3cm]
{\displaystyle } &  & 
  {k_{1}}_{0}=-\frac{1}{2\sqrt{S}}\Bigl(\frac{t}{x_{1}}+\frac{u}{x_{2}}\Bigr),\ 
  {k_{2}}_{0}=\frac{1}{2\sqrt{S}}\Bigl(\frac{s+t-z_{1}}{x_{1}}+\frac{s+u-u_{1}}{x_{2}}\Bigr).
\label{uie}\end{eqnarray}

The physical region $\Omega$ is determined by $\theta(R_{u_{1}})$,
where $-R_{u_{1}}$ is the Gram determinant, which has the form 
\begin{eqnarray}
 R_{u_{1}} = &&  \!\!\!\!\!\!\!\!\!  -A_{u_{1}}u_{1}^{2}-2B_{u_{1}}u_{1}-C_{u_{1}},\ \  A_{u_{1}}  =  -4m^{2}s+(s-v)^{2},
\nonumber \\[0.3cm]
B_{u_{1}} = &&  \!\!\!\!\!\!\!\!\! v[m^{2}(3s-v)+(s-v)(m_{q}^{2}-s-t+v)]+z[m^{2}(s-v)-m_{q}^{2}(s+v)+st+v(s+t-v)],
\nonumber \\[0.3cm]
C_{u_{1}} = &&  \!\!\!\!\!\!\!\!\! z^{2}[(m^{4}+m_{q}^{4}-2m^{2}(m_{q}^{2}+t-v)-2m_{q}^{2}(t+v)+(t-v)^{2}]+
\nonumber \\[0.3cm]
+ &&  \!\!\!\!\!\!\!\!\! 2zv[m^{4}+m_{q}^{4}+m_{q}^{2}(s-2t)-m^{2}(2m_{q}^{2}+s+2t-2v)+(t-v)(s+t-v)]+
\nonumber \\[0.3cm]
+ &&  \!\!\!\!\!\!\!\!\! v^{2}[m^{4}-2m^{2}(m_{q}^{2}+s+t-v)+(m_{q}^{2}-s-t+v)^{2}].
\end{eqnarray}

Then the total bremsstrahlung cross section has the form 
\begin{eqnarray}
 &  & \sigma_{R}=\frac{\alpha^{3}}{3}I_{\Omega}\bigl[T\bigr],\ \
T=\ \frac{1}{s^{2}}\sum\limits _{\chi=+,-}\sum\limits _{q=u,d,...}
\sum\limits _{i,j=\gamma,Z}{\lambda_{q}}_{\chi}^{i,j}{\lambda_{l}}_{\chi}^{i,j}\times
\nonumber \\[0.3cm]
{\displaystyle } &  & 
\Biggl([f_{q}^{A}(x_{1},Q^{2})f_{\bar{q}}^{B}(x_{2},Q^{2})+\chi f_{\bar{q}}^{A}(x_{1},Q^{2})f_{q}^{B}(x_{2},Q^{2})]
 [ Q_{q}^{2}{R_{qk}}_{\chi}^{q\bar{q}}\Pi^{i}{\Pi^{j}}^{*} + Q_{l}^{2}{R_{l}}_{\chi}^{q\bar{q}}D^{i}{D^{j}}^{*} ]
\nonumber \\[0.3cm]
{\displaystyle } &  & 
+[f_{q}^{A}(x_{1},Q^{2})f_{\bar{q}}^{B}(x_{2},Q^{2})-\chi f_{\bar{q}}^{A}(x_{1},Q^{2})f_{q}^{B}(x_{2},Q^{2})]
Q_{l}Q_{q}{R_{int}}_{\chi}^{q\bar{q}}\frac{\Pi^{i}{D^{j}}^{*}+D^{i}{\Pi^{j}}^{*}}{2}\Biggr).
\label{xshard}
\end{eqnarray}
Subscripts at $R$ (they can be found in Appendix A of \cite{yaf08}) indicate the origin of the emitted particle:
$qk$ -- quark for ISR both for photon and gluon [taking into account (\ref{zamena})],  
$l$ and $int$ --  lepton and interference term only for photon, respectively.
The boson propagators corresponding to the radiative case look like
\begin{equation}
\Pi^{j}=\frac{1}{s-z-v-m_{j}^{2}+im_{j}\Gamma_{j}}.
\end{equation}

We use the standard (noncovariant) method of IR singularity
separation dissecting the region of integration with the help of the
function $\theta_{\omega}=\theta(\frac{v+z}{2\sqrt{s}}-\omega) $
 and dividing the cross section (\ref{xshard}) into two parts: the
first one corresponds to soft photons (gluons) with energy less then $\omega$
(it goes to IR singularity cancellation in formula (\ref{soft}) ) 
and the second one corresponds to hard ones with energy larger than $\omega$. 

\begin{figure*}
\vspace{-10mm}
\hspace*{5mm}
\scalebox{0.14}{\includegraphics{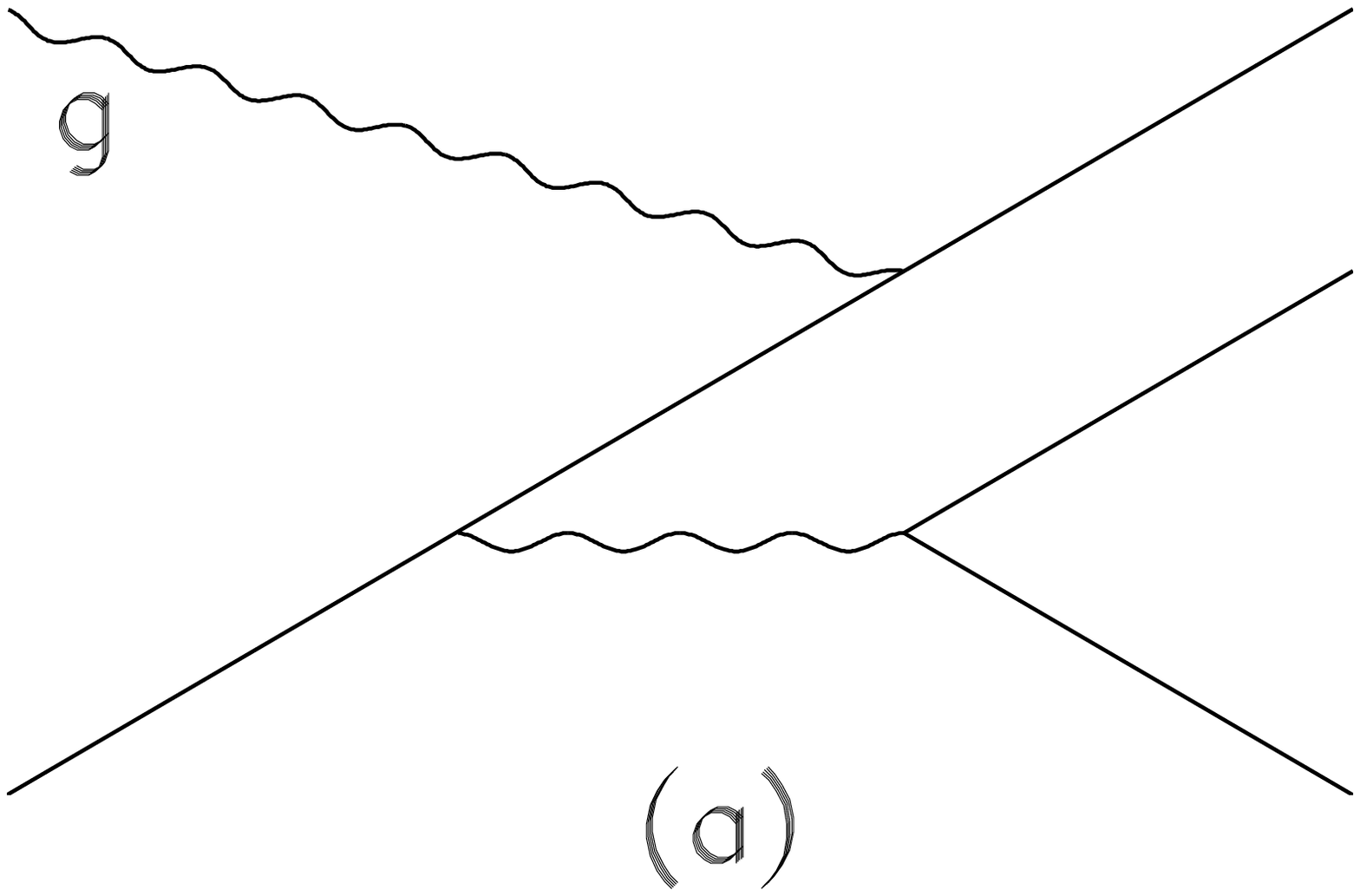}} \hspace*{3.5mm}
\scalebox{0.14}{\includegraphics{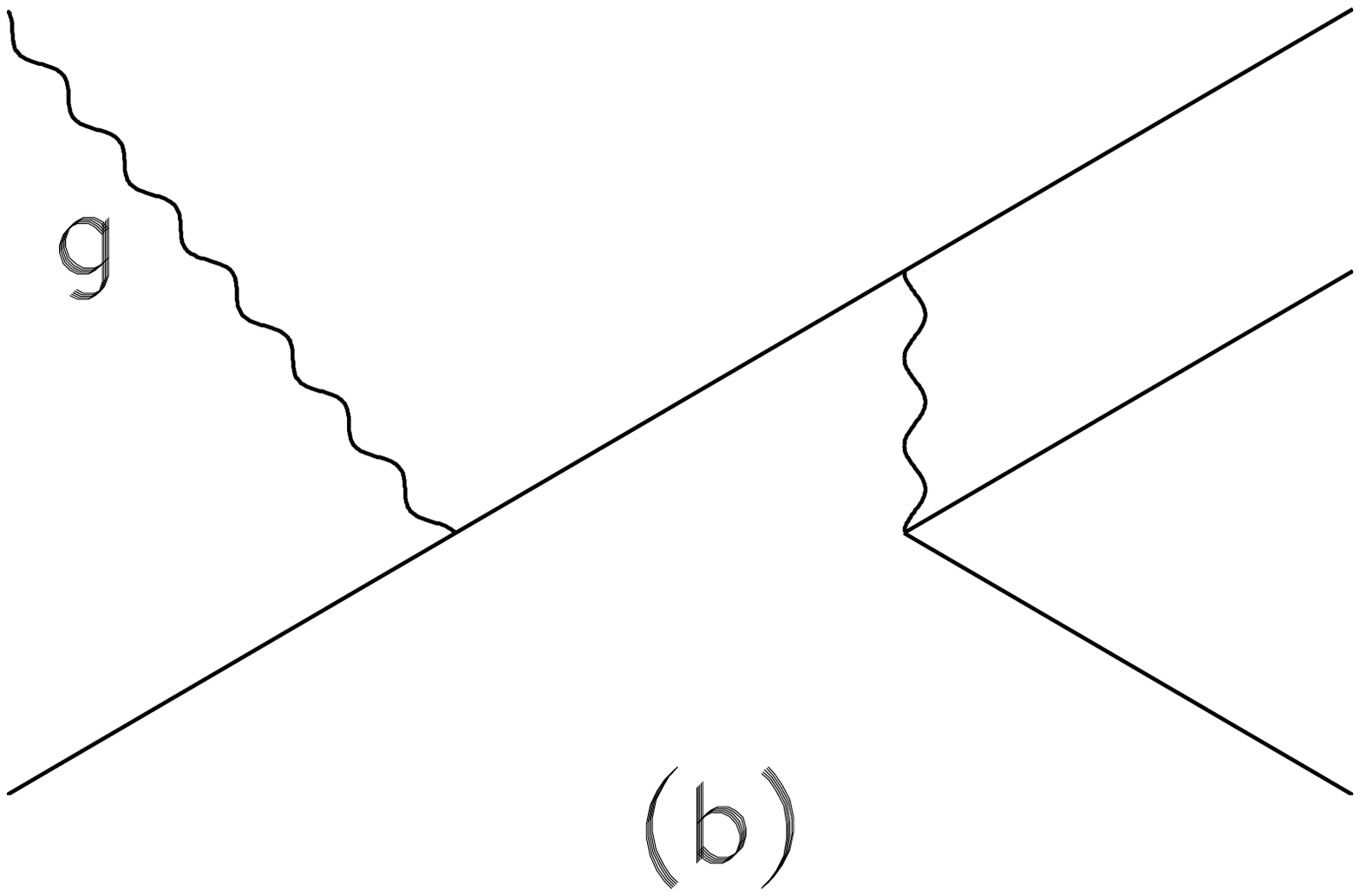}} \hspace*{3.5mm}
\scalebox{0.14}{\includegraphics{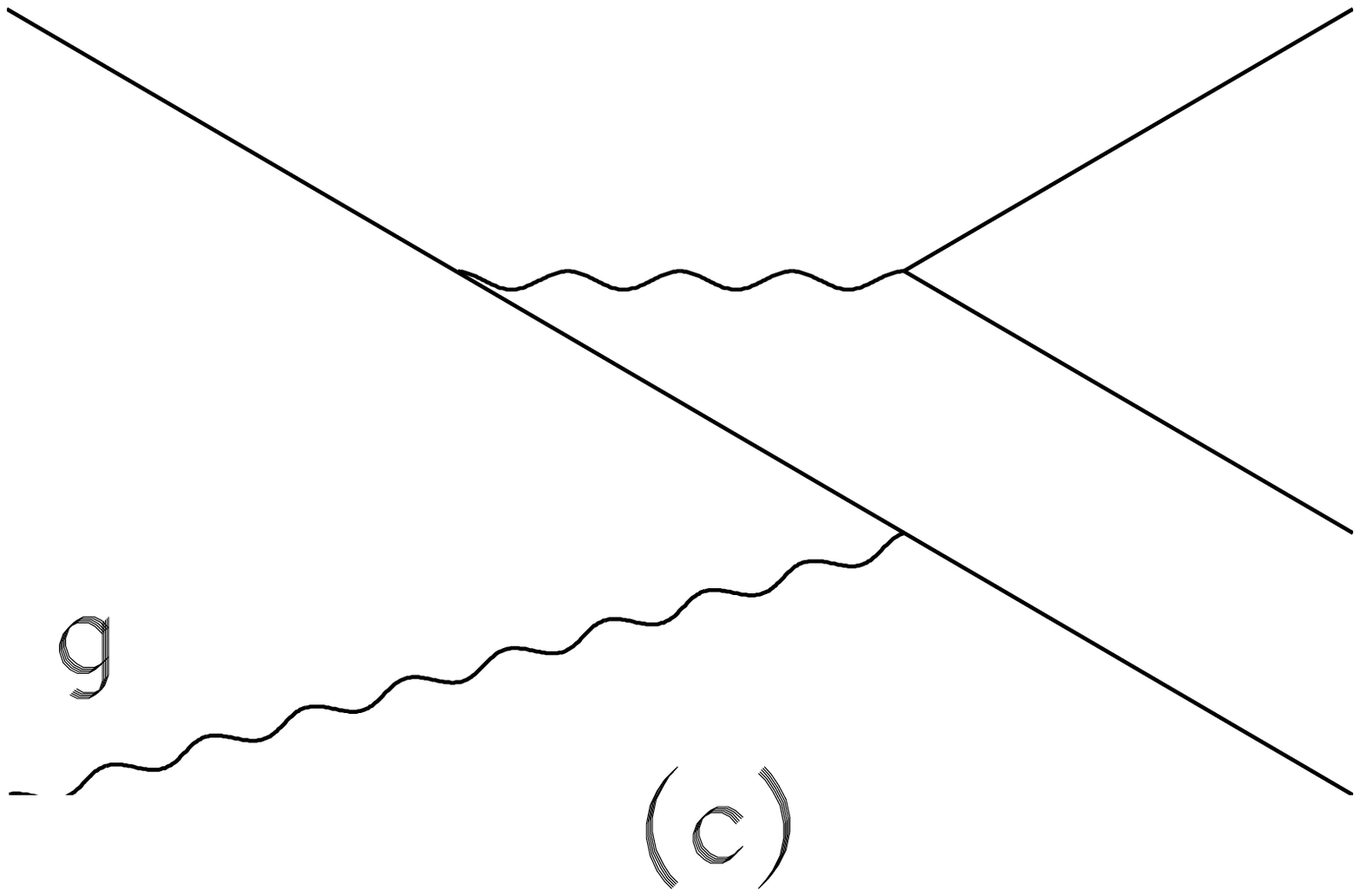}} \hspace*{3.5mm}
\scalebox{0.14}{\includegraphics{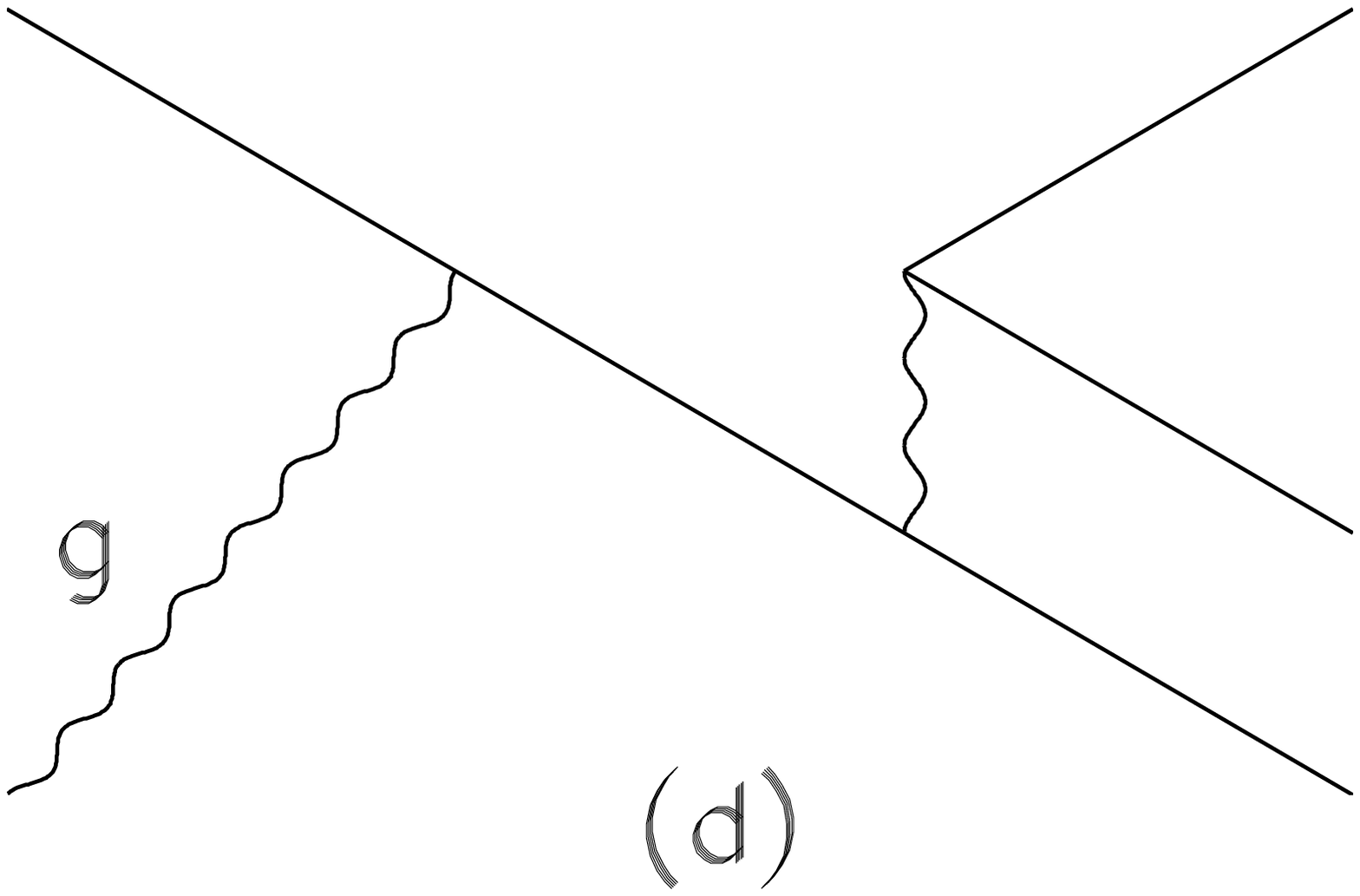}}
\vspace*{-2mm}
\caption{\label{fey-IGE}
Feynman graphs of $gq$-type (a, b) and $qg$-type (c, d)
for inverse gluon bremsstrahlung.
The unlabeled wavy lines stand for the virtual $\gamma$ or $Z$ boson. 
}
\end{figure*}

To finalize the calculation we have to take into consideration 
the inverse gluon emission (see, Fig.~\ref{fey-IGE}).
Methods of calculation are similar to cited above 
nonsinglet-channel (according to the QCD terminology) $q \bar q$-contributions.
All formulas for cross sections and  kinematics can be found in \cite{qcd2}.

\section{Fully differential cross section}

Here we rebuild all of the cross sections to fully differential form
\begin{equation}
\sigma_{C}\rightarrow\sigma_{C}^{(3)}\equiv\frac{d^{3}\sigma_{C}}{dMdyd\psi},\end{equation}
 where $y$ is the dilepton rapidity $y\equiv y(l^{+}l^{-})$, $\sigma_{C}^{(3)}$
are the corresponding contributions (or sum of contributions) of the
differential cross section, for example, if $C=0$, $\sigma_{0}^{(3)}$
is the differential Born cross section and so on.

For the part of the cross section with non-radiated kinematics, the
translation to differential form is easy to do using the Jacobian $J_{N}$: 
\begin{equation}
dx_{1}dx_{2}dt=J_{N}dMdyd\psi,\ \ J_{N}=\frac{4M^{3}e^{2y}}{S[1+\psi+(1-\psi)e^{2y}]^{2}}.
\end{equation}
 Then we get the following correlations: 
\begin{equation}
t=-\frac{M^{2}(\psi+1)}{1+\psi+(1-\psi)e^{2y}},\ x_{1}=e^{y}\frac{M}{\sqrt{S}},\ x_{2}=e^{-y}\frac{M}{\sqrt{S}},
\end{equation}
 and the differential cross section with non-radiative kinematics
looks like 
\begin{eqnarray}
\sigma_{N}^{(3)}=\frac{1}{3}J_{N}\sum\limits _{q=u,d,...}[f_{q}^{A}(x_{1},Q^{2})f_{\bar{q}}^{B}(x_{2},Q^{2})\sigma_{N}^{q\bar{q}}(t)+f_{\bar{q}}^{A}(x_{1},Q^{2})f_{q}^{B}(x_{2},Q^{2})\sigma_{N}^{\bar{q}q}(t)]\theta_{D}.
\label{xsfin-d}
\end{eqnarray}

For the radiative case, we rebuild the cross section to fully differential
form in a way analogical to the non-radiative case: 
\begin{equation}
dx_{1}dx_{2}dt=J_{R}^{(3)}dMdyd\psi,
\end{equation}
 where we use the correlations 
\begin{equation}
t=\frac{(u_{1}+M^{2})(e^{2y}(\psi-1)(v-u_{1})+(1+\psi)(z_{1}+M^{2}))}{e^{2y}(\psi-1)(u_{1}+M^{2})-(1+\psi)(z_{1}+M^{2})},
\label{t-rad}
\end{equation}
 \begin{equation}
x_{1}=\frac{e^{y}\sqrt{(u_{1}+M^{2})(v+z+M^{2})}}{\sqrt{S}\sqrt{z_{1}+M^{2}}},\ \ 
x_{2}=\frac{e^{-y}\sqrt{(z_{1}+M^{2})(v+z+M^{2})}}{\sqrt{S}\sqrt{u_{1}+M^{2}}}.
\label{x12-rad}
\end{equation}
 These are obtained from: 1) the formula for the radiative $s$: $s=M^{2}+v+z$, 
2) the formula for the radiative $\psi$ (\ref{psss}), 
3) the definition of dilepton rapidity $y$: 
$e^{2y}= (x_{1}/x_{2})(z_{1}+M^{2})/(u_{1}+M^{2})$.
Then, the Jacobian in radiative case can be expressed as 
\begin{equation}
J_{R}^{(3)}=\frac{4Me^{2y}}{S}\frac{(v+M^{2})(z_{1}+M^{2})(u_{1}+M^{2})}{[(1+\psi)(z_{1}+M^{2})+(1-\psi)e^{2y}(u_{1}+M^{2})]^{2}},
\end{equation}
 and satisfies $\lim_{k\to 0}J_{R}^{(3)}=J_{N}$. 
The differential
cross section corresponding hard bremsstralung is now given by 
\begin{eqnarray}
\sigma_{R}^{(3)}=\frac{\alpha^{3}}{3}\int\!\!\!\!\int\!\!\!\!\int\frac{1}{\pi\sqrt{R_{u_{1}}}}
\theta_{D}^{R}T\theta_{\omega}J_{R}^{(3)}dvdzdu_{1}.
\label{xshard-3}\end{eqnarray}

The remaining triple integral over the
physical bremsstrahlung region $\int \!\!\!\int\!\!\!\int ... dvdzdu_{1}$  
has to be computed numerically due to the complexity of the integration region and form of the integrand,
and due to the presence in the integrand of the intricate PDF,
which are $z,\ u_{1}$-dependent (see (\ref{x12-rad})). 
We can realize this numerical integration by Monte Carlo routine based on the VEGAS
algorithm \cite{VEGAS}, or simplify the hard bremsstrahlung contribution extracting 
the leading logarithm part and integrating $\int\!\!\!\int...dzdu_{1}$ analytically.
For further estimations we choose the last option. 
Exact formulas for QED CL parts can be found in \cite{LL},
results for nonsinglet QCD and singlet IGE CL parts can be found in \cite{qcd1} and \cite{qcd2}, respectively.

At last, starting with the fully differential cross sections, we can
construct the distributions over  $y$ and (or) $M$:
\begin{eqnarray}
\frac{d\sigma_{C}}{dMdy}= \int\limits _{-\zeta^{*}}^{\zeta^{*}}d\psi \sigma_{C}^{(3)}\theta_{D};\ \
\frac{d\sigma_{C}}{dM}= \int\limits _{-\zeta^{*}}^{\zeta^{*}}d\psi\int\limits _{y_{-}}^{y_{+}}dy\ \sigma_{C}^{(3)}\theta_{D},\ 
y_{\pm}=\pm\ln\frac{\sqrt{S}}{M}.
\label{di2}\end{eqnarray}

\section{Independence from unphysical parameters}

The proof of independence of the results from the parameter $\omega$
is rather simple and can be done numerically or analytically (see, for example, \cite{yaf08,LL}). 
For the soft-hard photon separator
we use $\omega=0.1$ GeV; however the results presented below do not
depend on $\omega$ in a wide interval: $1\ \mbox{GeV}\leq\omega\leq0.0001\ \mbox{GeV}$.

In order to solve the problem of quark mass singularity (QS), we used
the ${\rm \overline{MS}}$ scheme \cite{MSbar}, as in paper \cite{SANCz}.
After all of the prescribed manipulations, the part of the cross section
that must be subtracted in order to avoid the dependence on the quark mass assumes the form 
\begin{eqnarray}
\sigma_{{\rm QED (QCD), QS}}=\frac{1}{3} 
&  &  \!\!\!\!\!\!\!\!\!  \int\limits _{0}^{1}dx_{1}\int\limits _{0}^{1}dx_{2}
\int\limits _{-S}^{0}dt\int\limits _{0}^{1-{2\omega}/{M}}d\eta
\sum\limits _{q=u,d,...} \Bigl[\ \Bigl(q(x_{1})\Delta\bar{q}(x_{2},\eta)+
\nonumber \\[0.3cm]
{\displaystyle } &  & +\Delta q(x_{1},\eta)\bar{q}(x_{2})\Bigr)\sigma_{0}^{q\bar{q}}(t)
+(q\leftrightarrow\bar{q})\ \Bigr] \theta(s+t)\theta_{M}\theta_{D},
\label{quarksing}
\end{eqnarray}
\begin{eqnarray}
\!\!\!\!\!\Delta q(x,\eta)=\frac{1}{2} C_{\rm QED (QCD)}
\left[\frac{1}{\eta}q(\frac{x}{\eta},M_{{\rm sc}}^{2})\theta(\eta-x)-q(x,M_{{\rm sc}}^{2})\right]
\frac{1+\eta^{2}}{1-\eta}\left(\ln\frac{M_{{\rm sc}}^{2}}{m_{q}^{2}}-2\ln(1-\eta)-1\right),
\label{deq}
\end{eqnarray}
where $q(x) \equiv f_{q}(x,Q^{2})$, 
 and $M_{\rm sc}$ is the factorization scale \cite{MSbar}, which should be equal to $Q$ \cite{LL}. 
For the quark masses we used $m_{q}=m_{u}$,
although our numerical results practically do not depend on $m_{q}$ within the
interval $0.01m_{u}\leq m_{q}\leq 10m_{u}$. 
For IGE the result of QS-term substraction is trivial: 
$$\sigma_{\rm IGE}^{(3)} - \sigma_{\rm IGE, \rm QS}^{(3)}  = \sigma_{\rm IGE}^{(3)} (m_q \rightarrow M_{\rm sc}).$$

\section{Discussion of numerical results}

We investigate the scale of EWK and QCD corrections and their effect on the differential
observables of the Drell-Yan processes for CMS experiment using the FORTRAN program READY 
(Radiative corrEctions to lArge invariant mass Drell-Yan process) 
with the following set of parameters and prescriptions: 
\begin{itemize}
\item 
SM input electroweak parameters: 
$\alpha=1/137.035999679$,\ $m_{W}=80.398\ \mbox{GeV}$,\ $m_{Z}=91.1876\ \mbox{GeV}$,\ $\Gamma_{W}=2.141\ \mbox{GeV}$,\
$\Gamma_{Z}=2.4952\ \mbox{GeV}$, $m_{H}=125.7 \ \mbox{GeV}$;
\item 
muon mass $m_{\mu}=0.105658367\ \mbox{GeV}$, masses of the other
fermions for loop contributions to the BSE: $m_{e}=0.51099891\ \mbox{keV}$,\ $m_{\tau}=1.77699\ \mbox{GeV}$,\
$m_{u}=0.06983\ \mbox{GeV}$,\ $m_{c}=1.2\ \mbox{GeV}$,\ $m_{t}=174\ \mbox{GeV}$,
$m_{d}=0.06984\ \mbox{GeV}$,\ $m_{s}=0.15\ \mbox{GeV}$,\ $m_{b}=4.6\ \mbox{GeV}$;
(the light quark masses provide $\Delta \alpha_{had}^{(5)}(m_Z^2)$=0.0276);
\item 
modern MSTW2008 set of PDF \cite{MSTW2008}
with the choice $Q=M$;
\item 
taking into account 5 flavors of valence and sea quarks in the proton
(with the exception of the $t$ flavor) and set their masses as regulators
of the collinear singularity to $m_{q}=m_{u}$;
\item 
using "bare" setup for leptons identification requirements (no smearing, no recombination of lepton and photon). 
\end{itemize}

In Table 1 as example of READY output we show the relative corrections (RC) to Born differential cross section
\begin{equation}
\delta_{\rm C} = {\sigma^{(3)}_{\rm C}}/{\sigma^{(3)}_0}
\end{equation}
via different $y$, $\psi$ and $M$ 
with the muons in the final state ($l=\mu$), 
and the energy $\sqrt{S}=14\ \mbox{TeV}$ planned at the LHC in 2015.
\ \\
\ \\
{\bf Table 1. Relative corrections $\delta_{\rm NLO}$ via different $y$, $\psi$ and $M$.
}
{\small
\begin{center}
\begin{tabular}{|c|c||c||c||c|} 
\hline 
  \multicolumn{1}{|c|}{ $y$ }  & \multicolumn{1}{c||}{ $\psi$ } 
  & \multicolumn{1}{c||}{ $\delta_{\rm NLO}$ at $M$=1 TeV } 
  & \multicolumn{1}{c||}{ $\delta_{\rm NLO}$ at $M$=3 TeV  } 
  & \multicolumn{1}{c|}{ $\delta_{\rm NLO}$ at $M$=5 TeV  } \\
\hline                                                                                                                                                                
 0.0  &$ -0.8 $&$   -0.035 + 0.320 -0.134 $&$ -0.191 + 0.442 -0.068 $&$ -0.329 + 0.625 -0.054$\\
 0.0  &$ -0.4 $&$   -0.043 + 0.320 -0.089 $&$ -0.171 + 0.442 -0.061 $&$ -0.264 + 0.625 -0.051$\\
 0.0  &$  0.0 $&$   -0.036 + 0.320 -0.073 $&$ -0.154 + 0.442 -0.059 $&$ -0.231 + 0.625 -0.051$\\
 0.0  &$  0.4 $&$   -0.043 + 0.320 -0.089 $&$ -0.171 + 0.442 -0.061 $&$ -0.264 + 0.625 -0.051$\\
 0.0  &$  0.8 $&$   -0.035 + 0.320 -0.134 $&$ -0.191 + 0.442 -0.068 $&$ -0.329 + 0.625 -0.054$\\
\hline                                                                                                                                                              
 0.6  &$ -0.8 $&$  ~~0.008 + 0.456 -0.267 $&$ -0.146 + 0.582 -0.145 $&$ -0.101 + 0.770 -0.163$\\
 0.6  &$ -0.4 $&$   -0.014 + 0.453 -0.196 $&$ -0.138 + 0.577 -0.114 $&$ -0.114 + 0.773 -0.123$\\
 0.6  &$  0.0 $&$   -0.024 + 0.453 -0.140 $&$ -0.128 + 0.569 -0.090 $&$ -0.142 + 0.767 -0.090$\\
 0.6  &$  0.4 $&$   -0.028 + 0.443 -0.078 $&$ -0.145 + 0.565 -0.064 $&$ -0.223 + 0.759 -0.061$\\
 0.6  &$  0.8 $&$   -0.052 + 0.429 -0.059 $&$ -0.207 + 0.555 -0.052 $&$ -0.335 + 0.754 -0.049$\\
\hline                                                                                                                                     
 1.2  &$  0.0 $&$  ~~0.003 + 0.595 -0.295 $&$ -0.015 + 0.849 -0.215 $\\
 1.2  &$  0.4 $&$  ~~0.009 + 0.596 -0.205 $&$ -0.045 + 0.850 -0.147 $\\
 1.2  &$  0.8 $&$   -0.024 + 0.577 -0.047 $&$ -0.183 + 0.850 -0.056 $\\
\cline{1-4}
\end{tabular}
\end{center}
}
\ \\

Numbers presented in 3rd, 4th and 5th columns correspond to sum of all NLO contributions: 
NLO = EWK+QCD($q\bar q$)+QCD($qg$). 
Results for all RCs strongly depend on kinematical position, 
the necessary symmetry for different contributions is conserved.
Using different PDFs  (CTEQ6, MRST2004, MSTW2008) we did not mark any significant effect
for RCs in the whole kinematical region of CMS.

Let us now compare, as example, our EWK results with the numbers of several leading
world groups HORACE, SANC and ZGRAD presented in \cite{0803}. 
All parameters and detector conditions here are taken to be the same as in \cite{0803}. 
Our results for the relative correction to $d\sigma/dM$
at the point $M=1$~TeV ($l=\mu$, $\sqrt{S}$=14 TeV) 
is different by $\sim 1.5$\% comparing with HORACE and SANC. 
At $M=2$~TeV, this difference is $\sim 0.5$\%. The numbers of
the ZGRAD group in the region $0.9\ \mbox{TeV}\leq M\leq 1.8\ \mbox{TeV}$
are larger and are in better agreement with ours. We find such agreement
to be satisfactory, because, for the weak component of corrections,
we use the asymptotic approach \cite{PRD}, which greatly simplifies
the formulas and accelerates the calculation, but only works well
in the region $M > 0.5~\mbox{TeV}$, this explains why the agreement
becomes better with increasing $M$.

\section{Conclusions}

The complete NLO EWK and QCD radiative corrections
to the Drell-Yan process at large invariant dilepton  mass is studied in fully differential form.
The results for weak, QED, QCD parts are {the compact expressions}, 
they expand in Sudakov and collinear logarithms.
Using the FORTRAN code READY, the numerical analysis is performed
in the high-energy region corresponding to the CMS experiment at the CERN LHC. 
Both EWK and QCD RCs are found to become large at high dilepton mass
$M$ and to have the same order of magnitude as the systematic uncertainty
expected on CMS \cite{TDR}. 
Such large scale of RC does not allow neglecting the radiative correction procedure in the future experiments
on the Drell-Yan process with high dimuon masses at CMS LHC.
The exact NNLO QCD and ${\cal O}(\alpha\alpha_s)$ 
(see, for example \cite{Kataev}, where one of first understanding of role of such effects at high energies
had been achieved) corrections would be desirable for a better control of theory vs. experiment.

\section{Acknowledgments}

I thank local organizing committee of ACAT2013 and personally Prof. Jian-Xiong Wang and Bin Gong 
for help, finacial support and, as result, a happy possibility to take part in conference.
I am grateful to Prof. A. L. Kataev for the interest to the work  and support.
I would like to thank A.~Aleksejevs, A.~Arbuzov, S.~Barkanova, E.~Dydyshko,
E.~Kuraev, A.~Lanyov, S.~Pozzorini, V.~Mossolov, S.~Shmatov  and  N.~Shumeiko for the stimulating discussions.
I am grateful to A.~Arbuzov, S.~Bondarenko and D.~Wackeroth for a detailed comparison of part of the results.
I thank CERN (CMS Group), where part of this work was carried out, for warm hospitality during my visits.
Part of this work was supported by Belarus scientific program "Convergence".


\section*{References}
\begin{thebibliography}{30}
	\bibitem {new-boson} 
ATLAS Collaboration, Phys. Lett. B   716, 1  (2012);
CMS Collaboration,   Phys. Lett. B   716, 30 (2012).
	\bibitem{extra-bos} 
A.~Leike, Phys. Rep. \textbf{317}, 143 (1999). 
	\bibitem{extra-dim} 
N.~Arkani-Hamed \textit{et al.}, Phys. Lett. B \textbf{429}, 263 (1998); 
I.~Antoniadis \textit{et al.}, Phys. Lett. B \textbf{436}, 257 (1998);
L.~Randall and R.~Sundrum, Phys. Rev. Lett. \textbf{83}, 3370 (1999); 
4690 (1999) {[}hep-th/9906064{]}; 
C.~Kokorelis, Nucl. Phys. B \textbf{677}, 115 (2004). 
	\bibitem{cmsnote} 
I.~Belotelov \textit{et al.}, CERN-CMS-NOTE-2006-123. 
	\bibitem{FEWZ} 
Ye Li, Frank Petriello,  Phys. Rev. D \textbf{86} (2012) 094034.
	\bibitem{POWHEG} 
Luca Barze' \textit{et al.}, CERN-PH-TH-2013-027, arXiv:1302.4606 [hep-ph].
	\bibitem{CMS-PAS-EWK-11-007}
CMS Collaboration, CMS-PAS-EWK-11-007
	\bibitem{sud-log} 
V.~Sudakov, Sov. Phys. JETP \textbf{3}, 65 (1956). 
	\bibitem{YAFDY} 
V.~A.~Zykunov, Yad. Fiz. \textbf{69}, 1557 (2006)
(Engl. vers.: Phys. of Atom. Nucl. \textbf{69}, 1522 (2006)). 
	\bibitem{PRD} 
V.~A.~Zykunov, Phys. Rev. D \textbf{75}, 073019 (2007). 
	\bibitem{yaf08} 
V.~A.~Zykunov, Yad. Fiz. \textbf{71}, 757 (2008)
(Engl. vers.: Phys. of Atom. Nucl. \textbf{71}, 732 (2008)). 
	\bibitem{LL} 
V.~A.~Zykunov, Yad. Fiz. \textbf{73}, 1617 (2010)
(Engl. vers.: Phys. of Atom. Nucl. \textbf{73}, 1571 (2010)). 
	\bibitem{qcd1}
V.~A.~Zykunov, Yad. Fiz. \textbf{73}, 1269 (2010).
(Engl. vers.: Phys. of Atom. Nucl. \textbf{73}, 1229 (2010)). 
	\bibitem{qcd2}
V.~A.~Zykunov, Yad. Fiz. \textbf{74},  72 (2011).
(Engl. vers.: Phys. of Atom. Nucl. \textbf{74}, 72 (2011)). 
	\bibitem{BSH86} 
M.~B\"ohm, H.~Spiesberger and W.~Hollik, Fortschr. Phys. \textbf{34}, 687 (1986). 
	\bibitem{Hollik} 
W.~Hollik, Fortschr. Phys. \textbf{38}, 165 (1990). 
	\bibitem{baur} 
U.~Baur \textit{et al.}, Phys. Rev. D \textbf{57}, 199 (1998). 
	\bibitem{VEGAS} 
G.~Peter Lepage, J. Comput. Phys. \textbf{27}, 192 (1978). 
	\bibitem{MSbar} 
W.~A.~Bardeen \textit{et al.}, Phys. Rev. D \textbf{18}, 3998 (1978). 
	\bibitem{SANCz} 
A.~Arbuzov \textit{et al.}, Eur. Phys. J. C. \textbf{54}, 451 (2008). 
	\bibitem{MSTW2008} 
A.~D.~Martin \textit{et al.}, Eur. Phys. J. C \textbf{63}, 189 (2009). 
	\bibitem{0803} 
C.~Buttar \textit{et al.}, Proc. of Les Houches 2007,
Physics at TeV colliders, 121 p., arXiv:0803.0678 [hep-ph]. 
	\bibitem{TDR} 
CMS Physics TDR: V. II, Physics Performance, The CMS Collaboration, J. Phys. G. \textbf{34}, 995 (2007) 
	\bibitem{Kataev} 
A.~L.~Kataev, Phys. Lett. B \textbf{287}, 209 (1992) 




\end{thebibliography}
\end{document}